\documentclass[journal]{IEEEtran}

\usepackage{booktabs}
\usepackage{tabularx}
\usepackage{amssymb}
\usepackage{amsbsy}
\usepackage{amsmath}
\usepackage{bm}
\usepackage{verbatim}
\usepackage{mathrsfs}
\usepackage{amsfonts}
\usepackage{graphicx}
\usepackage[tight,footnotesize]{subfigure}
\usepackage[10pt]{moresize}
\usepackage{array}
\usepackage{color}
\usepackage{epsfig}
\usepackage{stfloats}
\usepackage{balance}

\usepackage{multirow}
\usepackage{makecell}
\usepackage{cancel}

\usepackage[noend]{algpseudocode}
\usepackage{algorithmicx,algorithm}

\usepackage{setspace}
\usepackage{cases}

\usepackage{epstopdf}
\usepackage{extarrows}

\usepackage{pifont} 
\usepackage{enumerate}
\usepackage{enumitem}

\newcommand{\subparagraph}{}
\usepackage{titlesec}
\usepackage[colorlinks]{hyperref}
\titlespacing{\section}{0pt}{2 ex plus .0ex minus .0ex}{1ex plus .0ex}
\titlespacing{\subsection}{0pt}{1.5 ex plus .0ex minus .0ex}{0.8 ex plus 0.0ex}
\titlespacing{\subsubsection}{0pt}{0.5ex plus .0ex minus .0ex}{0.0ex plus .0ex}

\allowdisplaybreaks[4]

\ifCLASSINFOpdf

\else

\fi

\allowdisplaybreaks[4]

\begin{document}

%\title{Uplink-Completion-Triggered GPU Execution for Low-Latency Multi-Agent Cooperative Perception}
%\title{Uplink-Completion-Triggered DNN Inference on an Edge GPU for Multi-Agent Cooperative Perception}
\title{Uplink-Completion-Triggered Edge-GPU Inference for Multi-Agent Cooperative Perception}

\author{Sai Xu, Yanan~Du, Chong Tang, and Gaojie~Chen
\vspace{-3mm} 

\thanks{S. Xu and C. Tang are with University College London, London, UK (e-mail: \texttt{sai.xu@ieee.org, chong.tang.18@ucl.ac.uk}). Y.~Du is with the Department of Electronic and Electrical Engineering, University of Sheffield, Sheffield, S1 4ET, UK (e-mail: \texttt{yanan.du@ieee.org}). G. Chen is with the School of Flexible Electronics (SoFE), Sun Yat-sen University, Shenzhen, Guangdong 518107, China (e-mail: \texttt{gaojie.chen@ieee.org}).}
}

\maketitle

\begin{abstract}
This paper investigates the coupling between wireless input completion and graphics processing unit (GPU) execution in centralized multi-agent cooperative perception. Specifically, beyond the conceptual treatment of completion-triggered overlap, a complete execution path is realized and validated on a physical GPU for a cooperative-perception deep neural network (DNN). Each encoder branch is released immediately upon completion of its corresponding input transmission, while the original fusion dependencies and inference mapping are preserved. The resulting release-triggered communication--computation coupling (RTCC) propagates validated wireless-completion events through host-to-device (H2D) staging, CUDA synchronization, and dependency-preserving branch dispatch, while remaining compatible with causal wireless schedulers. Experiments combining trace-driven wireless arrivals, physical-GPU execution, and measured-DAG evaluation show that RTCC reduces complete-detection latency across different communication loads and schedulers, while preserving identical detection outputs and average-precision performance relative to conventional execution.
\end{abstract}


\section{Introduction}\label{sec:introduction}

\IEEEPARstart{M}{ulti-agent} cooperative perception enables spatially distributed agents to integrate complementary observations, thereby extending perception beyond the capability of any individual agent \cite{Wang2020V2VNet,Hu2022Where2comm}. In a centralized edge-assisted architecture, agents upload local observations to an edge server, where deep neural network (DNN)-based fusion and inference generate a unified perception output. A widely adopted approach is intermediate fusion, in which agent-specific encoders extract intermediate features that are subsequently processed by shared fusion and task modules for joint inference \cite{Wang2020V2VNet,Hu2022Where2comm,Li2021DiscoNet}. This architecture naturally forms a fork--join execution structure, where parallel encoding branches converge at the shared downstream modules. Consequently, time-varying wireless arrivals are coupled with precedence-constrained DNN execution, making end-to-end latency reduction a joint communication and computation scheduling problem.

In cooperative perception, communication efficiency is primarily improved through collaborator selection, feature compression, and selective information exchange \cite{Hu2022Where2comm,Liu2020When2com,Qiu2022AutoCast,Liu2023AdaMap}. Meanwhile, cross-layer inference studies optimize task offloading, model partitioning, and the joint allocation of communication and computation resources \cite{Zaki2024QCPTO,Dong2025ITORA,Li2024ModelSplitting}, while graphics processing unit (GPU)-serving systems schedule DNN requests or operators after they become executable \cite{Gujarati2020Clockwork,Cui2022DVABatch,Ding2021IOS}. Collectively, these approaches determine what information to transmit, where computation should be performed, and how ready GPU workloads should be scheduled. However, they generally do not treat the completion of each mandatory wireless payload as the explicit release event for its corresponding branch in the fork--join DNN. This disconnect leaves the execution path from wireless delivery to host-to-device (H2D) transfer and GPU dispatch implicit, potentially causing ready encoder branches to remain idle while other inputs are still in transit.

This disconnect is commonly manifested as an all-arrival barrier in conventional implementations. Even when some agent inputs arrive much earlier than others, their corresponding encoder branches remain blocked until all transmissions are complete. As a result, the GPU performs no useful branch computation during the arrival gap and begins executing the encoders, fusion modules, and task heads only after the communication phase has fully ended. Such waiting is not imposed by the DNN itself, since the encoder branches are independent before the fusion join and can execute as soon as their own inputs become available. This observation creates an opportunity to overlap mandatory encoder computation with ongoing communication, without dropping inputs, approximating fusion, or altering the final inference result.

To realize this communication--computation overlap on an actual GPU execution path, this paper develops a physical-GPU realization of release-triggered communication--computation coupling (RTCC), in which each completed input transmission releases its corresponding encoder branch on the GPU. The emphasis is on realizing the complete communication-completion-to-GPU-release path and preserving the dependencies of a trained cooperative-perception network, rather than providing another simulation-only illustration of the release principle. Specifically, RTCC propagates the completion signal through the H2D transfer and CUDA dispatch path of a fully trained cooperative-perception DNN. Once an input is validated as complete, it is asynchronously staged to the GPU, after which a corresponding CUDA event releases its encoder branch on the compute stream. The resulting feature is written to a fixed agent-indexed location, while the shared fusion and task subgraph is launched only after all required features become available. Therefore, RTCC changes only the timing of mandatory computation, while preserving the original inputs, operators, dependencies, model parameters, and final inference mapping. The main contributions are summarized as follows:

\begin{itemize}[leftmargin=*]
\item A physical-GPU realization of RTCC is developed for fork--join cooperative-perception DNNs. Beyond prior simulation-level demonstrations, the proposed architecture connects completion events generated by trace-driven wireless emulation to pinned-memory H2D staging, CUDA-event synchronization, and encoder-branch dispatch on the target GPU, establishing a complete release path from communication completion to hardware-level DNN execution. The design remains compatible with causal communication schedulers that operate without future channel information.
\item A dependency-preserving execution design is established for asynchronous branch release. Through validated release control, profiled directed acyclic graph (DAG) construction, deterministic ready-task ordering, and fixed-index feature storage, the design ensures that each encoder branch is released only after its complete input becomes available and that out-of-order arrivals do not alter agent-to-feature correspondence. It advances eligible encoder computation while preserving all mandatory inputs, the original fusion dependencies, the all-agent join, and the trained inference mapping.
\item The latency benefit of RTCC is characterized analytically and validated against conventional all-arrival execution. A paired execution model is formulated to derive latency properties for precedence-constrained DAGs under the considered single-stream execution model, including the evaluated fork--join structure. Furthermore, physical-GPU experiments, trace-driven wireless emulation, resident-input CUDA measurements, measured-DAG studies, and output-invariance tests are conducted to evaluate latency reduction, analysis--measurement agreement, execution fidelity, and inference equivalence.
\end{itemize}

The remainder of this paper is organized as follows. Section~\ref{sec:related_work} reviews related studies on cooperative perception, edge inference, and GPU execution. Section~\ref{sec:system_model} presents the system model and problem formulation. Section~\ref{sec:method} describes the RTCC architecture, its dependency-preserving GPU realization, and the associated communication
policies. Section~\ref{sec:evaluation} reports the experimental methodology and results. Finally, Section~\ref{sec:conclusion} concludes the paper.

\section{Related Work}
\label{sec:related_work}

Related work spans three main directions: communication-efficient cooperative perception, cross-layer inference and resource allocation, and collaborative DNN inference and GPU execution. These studies optimize information exchange, computation placement, and ready-workload scheduling at different layers of the inference pipeline. However, the branch-level coupling between input completion and DNN execution remains insufficiently explored. Table~\ref{tab:related_work_comparison} summarizes the distinctions between RTCC and representative approaches.

\subsection{Communication-Efficient Cooperative Perception}
\label{subsec:rw_cooperative_perception}

Communication-efficient cooperative perception reduces collaboration overhead by adapting who communicates, what information is exchanged, and how shared features are aggregated. When2com learns communication partners through a handshake mechanism~\cite{Liu2020When2com}, while V2VNet performs iterative feature exchange over an agent graph~\cite{Wang2020V2VNet}. DiscoNet further learns spatially varying collaboration through teacher--student distillation~\cite{Li2021DiscoNet}. These methods improve efficiency by adapting the collaboration topology and feature-aggregation process.
More recent approaches reduce transmitted content at finer spatial, channel, or semantic granularity. Where2comm selects perceptually critical regions using spatial confidence maps~\cite{Hu2022Where2comm}, whereas How2comm combines spatial--channel filtering with temporal compensation for delayed features~\cite{Yang2023How2comm}. CodeFilling transmits compact codebook indices and reconstructs receiver-relevant information~\cite{Hu2024CodeFilling}, while CoSDH combines supply--demand-aware region selection with intermediate--late hybrid fusion~\cite{Xu2025CoSDH}. Their gains arise from selectively transmitting, compressing, reconstructing, or compensating collaborative information. Recent studies extend this direction to asynchronous, sparse, and heterogeneous collaboration. TraF-Align reconstructs current-time features from delayed observations to mitigate spatial and semantic misalignment, Long-SCOPE employs a fully sparse representation for long-range cooperative perception, and CodeAlign translates heterogeneous features through a shared codebook~\cite{Song2025TraFAlign,Wang2026LongSCOPE,Liu2026CodeAlign}. 

Overall, existing methods improve communication efficiency or robustness by modifying the collaboration pattern, exchanged representation, temporal alignment, or fusion process, while treating wireless delivery primarily as a bandwidth or delay constraint. In contrast, RTCC preserves all mandatory inputs and the original fork--join inference graph, and uses each physical payload completion as the release event for its corresponding encoder branch.

\subsection{Cooperative-Perception Systems and Cross-Layer Inference}
\label{subsec:rw_cp_systems}

System-oriented cooperative-perception studies coordinate perception decisions with wireless transmission to improve latency and task utility. AutoCast prioritizes safety-relevant objects and schedules their distributed dissemination, with its communication design evaluated on a radio testbed~\cite{Qiu2022AutoCast}. AdaMap combines object selection, point-cloud compression and reconstruction, and adaptive detection--tracking to control tail latency as the collaboration scale increases~\cite{Liu2023AdaMap}. These systems demonstrate the value of perception-aware communication, but achieve efficiency by selecting or transforming shared observations and adapting the perception pipeline.
Cross-layer inference studies further optimize task placement and communication--computation resource allocation. Q-CPTO selects and offloads perception-aggregation tasks according to predicted trajectories and shared regions of interest~\cite{Zaki2024QCPTO}, while ITORA jointly optimizes sensing-subregion assignment, task offloading, and communication and computation resources under coverage constraints~\cite{Dong2025ITORA}. Related split-inference methods jointly determine DNN partition points and edge resources~\cite{Li2024ModelSplitting}, or incorporate downstream allocation objectives into differentiable traffic prediction~\cite{Lyu2024ObjectiveDriven}. These formulations typically abstract computation as an aggregate workload or a sequential model partition.

Overall, these approaches determine what is processed, where computation occurs, and how resources are allocated. They do not generally model each mandatory input completion as the release condition of a specific branch in a multi-branch DNN. By contrast, RTCC fixes the inputs, trained model, and edge-GPU placement, and exposes branch-specific completion events to the execution layer.

\subsection{Collaborative DNN Inference and GPU Execution}
\label{subsec:rw_dnn_gpu}

Collaborative DNN inference primarily optimizes model placement and execution across devices. Neurosurgeon profiles layer-level communication and computation costs to select a device--cloud partition~\cite{Kang2017Neurosurgeon}. SPINN jointly adapts model partitioning and early-exit decisions to network conditions and latency objectives~\cite{Laskaridis2020SPINN}, while CoEdge distributes DNN workloads across heterogeneous edge devices according to their communication and computation capabilities~\cite{Zeng2021CoEdge}.
GPU-serving and graph-scheduling systems instead optimize workloads after they become ready. Clockwork coordinates low-level GPU actions for predictable multi-model inference~\cite{Gujarati2020Clockwork}. VELTAIR adapts compilation and scheduling for multi-tenant serving~\cite{Liu2022VELTAIR}, while DVABatch supports dynamic batching for multi-entry/multi-exit networks~\cite{Cui2022DVABatch}. IOS exploits inter-operator parallelism, and hardware-aware graph scheduling maps DNN subgraphs to accelerator resources~\cite{Ding2021IOS,Zhao2023GraphScheduling}. More recently, Torpor combines host-memory model residency, asynchronous GPU-runtime redirection, model swapping, and request scheduling for low-latency serverless inference~\cite{Yu2025Torpor}. These systems improve batching, operator ordering, and accelerator utilization, but generally treat workload readiness as an external condition.
Preliminary studies have used release-triggered execution mainly as a conceptual or simulation-level mechanism in other edge-inference settings. O-WiN studies wireless--accelerator pipelining under a multi-core neural-processor abstraction~\cite{Xu2026OWiN}, while a multi-UAV study evaluates sensing-data offloading with multi-branch edge inference primarily through scheduling-level simulation~\cite{Du2026MultiUAV}. Neither work realizes the input-completion signal through the pinned-memory H2D, CUDA-event, and branch-dispatch path of a complete cooperative-perception DNN on a physical GPU.

Overall, existing collaborative-inference and GPU-execution studies optimize model placement, ready-workload scheduling, or communication--accelerator coordination. In contrast, this work focuses on realizing and validating input-specific release through the actual H2D and CUDA execution path of a complete cooperative-perception DNN, including release validation, deterministic branch dispatch, fixed-index feature storage, and physical-GPU timing measurements, while preserving the original fork--join dependencies and inference mapping.

\begin{table*}[!t]
\centering
\caption{Comparison with representative related work.}
\label{tab:related_work_comparison}
\footnotesize
\setlength{\tabcolsep}{3pt}
\renewcommand{\arraystretch}{1.10}

\begin{tabularx}{\textwidth}{
@{}
p{0.19\textwidth}
X
>{\centering\arraybackslash}p{0.075\textwidth}
>{\centering\arraybackslash}p{0.075\textwidth}
>{\centering\arraybackslash}p{0.075\textwidth}
>{\centering\arraybackslash}p{0.075\textwidth}
@{}
}
\toprule
\textbf{Work} &
\textbf{Primary mechanism} &
\shortstack[t]{\textbf{Input}\\\textbf{fidelity}} &
\shortstack[t]{\textbf{DNN}\\\textbf{deps.}} &
\shortstack[t]{\textbf{Completion}\\\textbf{overlap}} &
\shortstack[t]{\textbf{GPU}\\\textbf{gating}} \\
\midrule

\multicolumn{6}{@{}l}{\textit{Communication-efficient cooperative perception}}\\

Where2comm~\cite{Hu2022Where2comm} &
Spatial feature selection and compression &
$\times$ & $\times$ & $\times$ & $\times$ \\

CodeFilling~\cite{Hu2024CodeFilling} &
Compact representation and reconstruction &
$\times$ & $\times$ & $\times$ & $\times$ \\

Recent CP methods
~\cite{Song2025TraFAlign,Wang2026LongSCOPE,Liu2026CodeAlign} &
Asynchronous alignment, sparse representation, and heterogeneous
feature translation &
$\times$ & $\times$ &
$\times$ & $\times$ \\

AutoCast/AdaMap~\cite{Qiu2022AutoCast,Liu2023AdaMap} &
Perception-aware dissemination &
$\times$ & $\times$ & $\times$ & $\times$ \\

\addlinespace[2pt]
\multicolumn{6}{@{}l}{\textit{Resource allocation and collaborative inference}}\\

Q-CPTO/ITORA~\cite{Zaki2024QCPTO,Dong2025ITORA} &
Task offloading and resource allocation &
$\times$ & $\times$ & $\times$ & $\times$ \\

SPINN~\cite{Laskaridis2020SPINN} &
Model partitioning and early exit &
$\times$ & $\checkmark$ & $\triangle$ & $\times$ \\

CoEdge~\cite{Zeng2021CoEdge} &
Distributed DNN partitioning &
$\checkmark$ & $\checkmark$ & $\triangle$ & $\times$ \\

\addlinespace[2pt]
\multicolumn{6}{@{}l}{\textit{DNN orchestration and GPU execution}}\\

IOS/DVABatch~\cite{Ding2021IOS,Cui2022DVABatch} &
Operator scheduling and batching &
$\checkmark$ & $\checkmark$ & $\times$ & $\times$ \\

Torpor~\cite{Yu2025Torpor} &
Low-latency serverless GPU inference &
$\checkmark$ & $\times$ &
$\times$ & $\times$ \\

O-WiN~\cite{Xu2026OWiN}$^{\dagger}$ &
Wireless--accelerator pipelining &
N/R & $\checkmark$ & $\checkmark$ & $\times$ \\

Multi-UAV scheduling~\cite{Du2026MultiUAV}$^{\dagger}$ &
Offloading and multi-branch scheduling &
N/R & $\checkmark$ & $\checkmark$ & $\times$ \\

\midrule
\textbf{RTCC} &
\textbf{Completion-triggered GPU execution} &
$\checkmark$ & $\checkmark$ & $\checkmark$ & $\checkmark$ \\
\bottomrule
\end{tabularx}

\vspace{0.5mm}
\parbox{0.98\textwidth}{\scriptsize
$\checkmark$: supported;
$\times$: not supported;
$\triangle$: partially supported;
N/R: not explicitly reported.
Input fidelity denotes preservation of all mandatory inputs and the complete
trained inference mapping. Completion overlap indicates that an input-completion
event can release its corresponding DNN computation. GPU gating requires
completion events to govern H2D staging, CUDA synchronization, and encoder
dispatch. $^{\dagger}$Preprint available at the time of writing.}
\end{table*}

\section{System Model}
\label{sec:system_model}

\begin{figure}[t]
    \centering
    \includegraphics[width=0.5\textwidth]{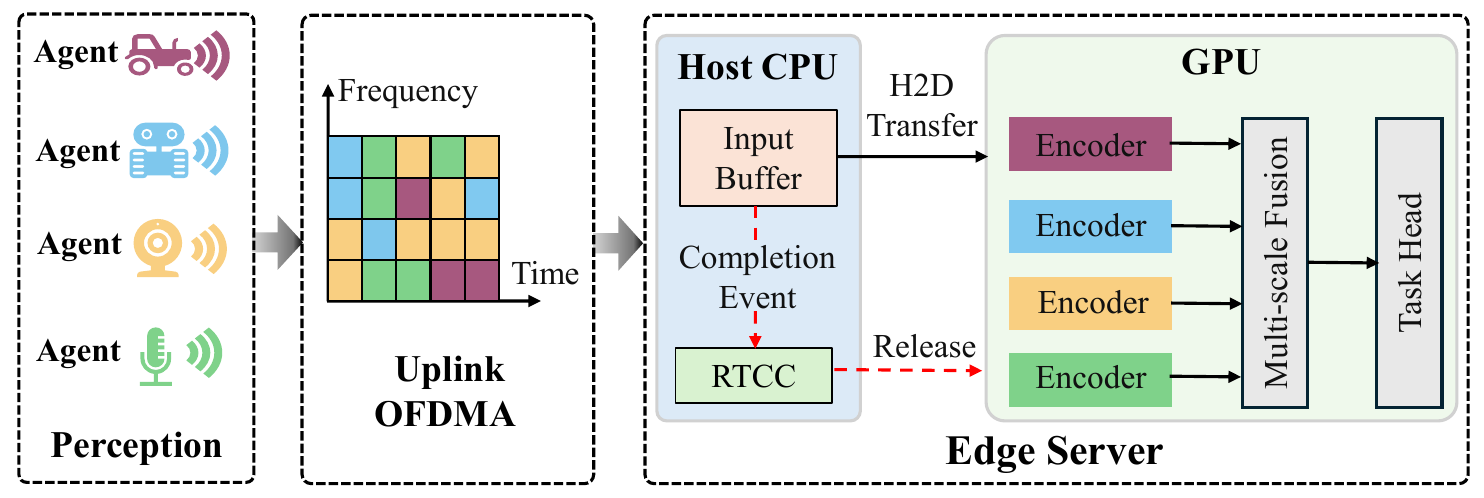}
    \caption{An illustration of the centralized multi-agent cooperative perception system.}
    \label{fig:system_model}
\end{figure}

Fig.~\ref{fig:system_model} illustrates a centralized multi-agent cooperative perception system consisting of a GPU-equipped edge server and $K$ networked sensing agents, where $\mathcal{K}\triangleq\{1,\ldots,K\}$ denotes the index set of sensing agents. For each synchronized perception frame, every agent generates a branch input and transmits it to the edge server over a shared wireless uplink. With a pre-trained multi-branch DNN deployed, the edge server leverages its GPU to process incoming inputs in three stages: independent encoding for each branch, multi-scale fusion of all encoded branches, and inference via the subsequent detection heads. The system objective is to minimize the end-to-end latency from the start of uplink transmission to the complete perception output, subject to communication causality, radio-resource feasibility, DNN precedence, and full participation of all branches.  

\subsection{Communication Model}
\label{subsec:communication_model}

The uplink transmission from the sensing agents to the edge server employs orthogonal frequency-division multiple access (OFDMA). Specifically, the available uplink resources are organized into a time-frequency grid, in which each grid element represents an orthogonal resource block (RB). Along the time dimension, the transmission interval is divided into slots indexed by \(\mathcal{T}\triangleq\{1,\ldots,T_{\max}\}\), each having a duration of \(\Delta\). Along the frequency dimension, the uplink bandwidth is partitioned into \(F\) orthogonal RBs indexed by \(\mathcal{F}\triangleq\{1,\ldots,F\}\). 

Let $x_{k,f}[t]\in\{0,1\}$ denote the RB-assignment indicator, where $x_{k,f}[t]=1$ if RB $f$ is allocated to agent $k$ in slot $t$, and $x_{k,f}[t]=0$ otherwise. The OFDMA orthogonality constraint is

\begin{equation}
    \sum_{k\in\mathcal{K}} x_{k,f}[t] \leq 1,
    \qquad \forall f\in\mathcal{F},\ t\in\mathcal{T},
    \label{eq:rb_orthogonality}
\end{equation}
which ensures that each RB is assigned to at most one sensing agent in each time slot. Meanwhile, a sensing agent may be allocated multiple RBs within the same slot. Let $R_{k,f}[t]$ denote the effective uplink transmission rate of sensing agent $k$ over RB $f$ in slot $t$. The amount of data transmitted by sensing agent $k$ during slot $t$ is
\begin{equation}
    d_k[t]
    = \Delta \sum_{f\in\mathcal{F}} x_{k,f}[t] R_{k,f}[t].
    \label{eq:slot_payload}
\end{equation}
The effective rate may incorporate propagation loss, fading, blockage, and frequency-selective channel variations.

Let $\ell_k$ denote the earliest slot in which the branch input of sensing agent $k$ becomes locally available for transmission. Accordingly, the agent cannot be allocated any RB before slot $\ell_k$, which is enforced by
\begin{equation}
    x_{k,f}[t] = 0,
    \qquad \forall k\in\mathcal{K},\ f\in\mathcal{F},\ t<\ell_k.
    \label{eq:local_ready_constraint}
\end{equation}
Let $B_k$ and $Q_k[t]$ denote the branch input size and the remaining undelivered payload of sensing agent $k$ at the start of slot $t$, respectively. With $Q_k[1]=B_k$, the queue evolves as
\begin{equation}
    Q_k[t+1]
    =
    \max\left\{0,\,Q_k[t]-d_k[t]\right\},
    \quad \forall k\in\mathcal{K},\ t\in\mathcal{T}.
    \label{eq:queue_evolution}
\end{equation}
Accordingly, the physical uplink completion time of branch $k$ is
\begin{equation}
    a_k
    =
    \Delta
    \min\left\{
        t\in\mathcal{T}:
        \sum_{\tau=1}^{t} d_k[\tau] \geq B_k
    \right\}.
    \label{eq:branch_arrival}
\end{equation}
If the set in~\eqref{eq:branch_arrival} is empty, $a_k=+\infty$ is set.

The uplink resource-allocation policy is restricted to be causal. Define
\begin{equation}
    \mathcal{H}_t
    \triangleq
    \left\{
        \mathbf{R}[1{:}t],
        \mathbf{Q}[1{:}t],
        \mathbf{x}[1{:}t-1],
        \mathbf{B},
        \boldsymbol{\ell}
    \right\}
    \label{eq:causal_history}
\end{equation}
as the information available prior to the RB allocation in slot $t$, where $\mathbf{R}[1{:}t]$, $\mathbf{Q}[1{:}t]$, and $\mathbf{x}[1{:}t-1]$ collect the observed uplink rates, queue states, and
past allocation decisions, respectively. A communication policy
$\pi^{\mathrm{C}}$ is causal if
\begin{equation}
    \mathbf{x}[t]
    =
    \pi^{\mathrm{C}}\!\left(\mathcal{H}_t\right),
    \label{eq:causal_policy}
\end{equation}
without access to future rate realizations $\mathbf{R}[t+1{:}T_{\max}]$. Different causal policies may yield different branch-arrival vectors. Nevertheless, the DNN computation depends on the branch-arrival vector $\mathbf{a}\triangleq[a_1,\ldots,a_K]^{\mathsf T}$, rather than on the policy used to generate it.

\subsection{DNN Computation Model}
\label{subsec:computation_model}

At the edge server, the DNN inference workload is represented by a profiled directed acyclic graph (DAG), denoted by $\mathcal{G}=(\mathcal{V},\mathcal{E})$, where each node $v\in\mathcal{V}$ represents a computational task. The communication-dependent prefix contains one encoder $E_k\in\mathcal{V}$ for each sensing agent $k\in\mathcal{K}$. These encoders are mutually independent and can be executed after their respective branch inputs arrive. The shared post-join subgraph comprises multi-scale fusion, upsampling, feature concatenation, and the subsequent classification and regression heads. Since each fusion branch depends on all $K$ encoders, encoder computation may overlap ongoing uplink transmission, whereas fusion can only start after all required encoders have completed execution. The structure of the considered DAG is illustrated in Section~\ref{subsubsec:dnn_gpu_platform}.

Each node $v\in\mathcal{V}$ has a nonnegative service time $c_v$, obtained by profiling the trained DNN on the target GPU. Let $s_v$ and $f_v$ denote the start and finish times of node $v$, respectively. They satisfy
\begin{equation}
    f_v = s_v + c_v,
    \label{eq:node_finish}
\end{equation}
and the DAG precedence constraints require
\begin{equation}
    s_v \geq f_u,
    \qquad \forall (u,v)\in\mathcal{E}.
    \label{eq:dag_precedence}
\end{equation}
The edge GPU is modeled as a single nonpreemptive, work-conserving compute stream, consistent with the profiled execution configuration. Hence, positive-duration DAG nodes do not overlap on the GPU, although encoder execution may overlap wireless transmission. This operator-level model captures the measured service and dependency structure relevant to scheduling, without treating the GPU as a set of independent, equal-speed processors. Additionally, each encoder $E_k$ is subject to
\begin{equation}
     s_{E_k} \geq r_k,
    \qquad \forall k\in\mathcal{K},
    \label{eq:encoder_release}
\end{equation}
where $r_k$ denotes the release time of encoder $E_k$. The release time specifies only the earliest feasible start of an encoder.

\subsection{Execution and Latency Model}
\label{subsec:execution_latency_model}

Let $\mathbf{r}\triangleq[r_1,\ldots,r_K]^{\mathsf T}$ denote the branch-release vector generated by a release rule $\rho:\mathbf{a}\mapsto\mathbf{r}$. Physical release causality requires
\begin{equation}
    r_k\geq a_k,
    \qquad \forall k\in\mathcal{K},
    \label{eq:release_causality}
\end{equation}
such that no edge-side encoder can be released before its full input has arrived. Together, \eqref{eq:encoder_release} and \eqref{eq:release_causality} imply $s_{E_k}\geq a_k$. The admissible set of release rules satisfying~\eqref{eq:release_causality} is denoted by $\Pi_{\mathrm{rel}}$. Let $o\in\mathcal{V}$ denote the output node after the task heads. The full communication-plus-inference latency is
\begin{equation}
    T\triangleq f_o,
    \label{eq:full_latency}
\end{equation}
where time is measured from the start of the frame's uplink transmission. To separate communication completion from the remaining edge computation, define the final branch-arrival time and the post-arrival GPU tail as
\begin{equation}
    A\triangleq\max_{k\in\mathcal{K}}a_k,
    \qquad
    T_{\mathrm{tail}}\triangleq T-A.
    \label{eq:gpu_tail}
\end{equation}
The task is complete only if every branch input is delivered and all nodes required by $o$ are executed. The release rule may change node start times but must not alter the DNN inputs, numerical operations, dependencies, or fixed-index fusion order.
 
\subsection{Problem Formulation}
\label{subsec:problem_formulation}

Let $\mathbf{a}_{\pi} \triangleq  \mathcal{A}\!\left(\pi^{\mathrm C}, \mathbf{R}, \mathbf{B}, \boldsymbol{\ell} \right)$ denote the branch-arrival vector induced by a causal communication policy
$\pi^{\mathrm C}$, where $\mathbf{R}$, $\mathbf{B}$, and $\boldsymbol{\ell}$ collect the uplink-rate realizations, branch-input sizes, and local availability-slot indices, respectively. Given a release rule $\rho\in\Pi_{\mathrm{rel}}$, define $\mathbf{r}_{\rho}\triangleq\rho(\mathbf{a}_{\pi})$. Furthermore, let $\mathcal{S}(\mathbf{a},\mathbf{r},\mathcal{G},\mathbf{c})$ denote the
completion time of full DNN inference given arrival vector $\mathbf{a}$, release vector $\mathbf{r}$, DNN DAG $\mathcal{G}$, and service-time vector $\mathbf{c}$. The mapping $\mathcal{S}$ incorporates the release constraints, DAG precedence constraints, and the single-stream GPU execution model introduced above.

The resulting end-to-end latency-minimization problem is
\begin{subequations}
\label{prob:rtcc_e2e}
\begin{align}
\mathrm{(P1)}\quad
\underset{\substack{\pi^{\mathrm C}\in\Pi_{\mathrm{causal}},\,
                     \rho\in\Pi_{\mathrm{rel}}}}
          {\min}
\; &
\mathbb{E}_{\mathbf{R}}\!\left[
    \mathcal{S}\!\left(
        \mathbf{a}_{\pi},
        \mathbf{r}_{\rho},
        \mathcal{G},
        \mathbf{c}
    \right)
\right]
\label{prob:rtcc_e2e_obj} \\
\mathrm{s.t.}\quad
& \eqref{eq:rb_orthogonality},
  \eqref{eq:local_ready_constraint},
  \eqref{eq:queue_evolution}, \notag\\[-0.5mm]
& \eqref{eq:dag_precedence},
  \eqref{eq:encoder_release},
  \eqref{eq:release_causality},\notag\\[-0.5mm]
& x_{k,f}[t]\in\{0,1\},\quad
  \substack{\forall k\in\mathcal K,\ f\in\mathcal F,\\ t\in\mathcal T},
\label{prob:binary_allocation} \\
& a_k<+\infty,\quad \forall k\in\mathcal K,
\label{prob:finite_arrival}
\end{align}
\end{subequations}
Here, $\Pi_{\mathrm{causal}}$ denotes the set of policies satisfying \eqref{eq:causal_history}--\eqref{eq:causal_policy}. The referenced constraints enforce OFDMA orthogonality, local-input availability, payload evolution, physical release causality, encoder eligibility, and DNN precedence. Constraint~\eqref{prob:binary_allocation} specifies the binary allocation domain, while \eqref{prob:finite_arrival} requires every branch to be delivered within the communication horizon. Nonoverlapping GPU execution is embedded in $\mathcal{S}$. Problem~(\text{P1}) serves as the joint design objective and does not imply that a globally optimal online solution is computed.

\section{Release-Triggered Communication--Computation Coupling}
\label{sec:method}

Problem~(\text{P1}) expresses the joint objective over a causal communication policy and an admissible release rule for minimizing the expected completion time of the entire cooperative perception pipeline. The key challenge is that each RB-allocation decision affects the branch-arrival vector $\mathbf a$, which in turn determines when the corresponding edge-side DNN branches become executable and how much computation can overlap the remaining uplink transmissions. This temporal coupling, together with time-varying channels and the absence of future CSI, makes exhaustive joint optimization of communication decisions and GPU execution schedules
impractical for online operation. 

To facilitate online operation, this section proposes release-triggered communication--computation coupling (RTCC), an event-driven architecture that decouples causal communication scheduling from DNN execution while preserving their coupling through physical input-completion events. RTCC comprises an offline profiling stage and an online event-driven execution
stage. The offline stage extracts the computational DAG from the trained perception DNN and profiles the branch-input sizes and node service times on the target GPU. The online stage integrates three modules: a causal communication policy, a release controller, and a dependency-aware GPU executor. 

\subsection{Offline DAG Construction and Profiling}
\label{subsec:offline_profiling}

Prior to online execution, the pretrained DNN is represented as a profiled DAG that captures its computational structure and execution dependencies on the target GPU. To construct this representation, the DNN is decomposed into \(K\) agent-specific branch encoders \(\{E_k\}_{k\in\mathcal{K}}\) and a shared post-encoder subgraph. The shared subgraph contains all subsequent operations, including multi-scale feature fusion, upsampling, tensor concatenation, and task-head inference. The tensor dependencies among these components define the edge set \(\mathcal{E}\) of \(\mathcal{G}=(\mathcal{V},\mathcal{E})\). In particular, each fusion stage retains all branch features required by the original network. Consequently, the resulting DAG preserves the complete inference dependencies without branch dropping, feature pruning, or approximate fusion.

In addition to extracting the graph structure, the offline stage profiles the computational and communication characteristics required for online scheduling. For each node \(v\in\mathcal{V}\), its nonpreemptive execution time is measured repeatedly on the target GPU using CUDA events after device warm-up, yielding a representative service time \(c_v\). The resulting service times form the vector \(\mathbf{c}=[c_v]_{v\in\mathcal{V}}\). Meanwhile, the input size for branch encoder \(E_k\) is measured and defined as the corresponding communication payload \(B_k\). The resulting DAG structure, service-time vector, and branch payload sizes jointly provide the offline information required for online communication scheduling and GPU execution. This profiling procedure is performed once for each DNN--hardware configuration and is independent of channel realizations and future CSI. During online inference, the profiled DAG is executed using the nonpreemptive single-stream GPU model defined in Section~\ref{subsec:computation_model}.

\subsection{Causal Communication and Completion Events}
\label{subsec:causal_event_generation}

At the beginning of slot \(t\), the communication module identifies the set of transmission-eligible agents as
\begin{equation}
    \mathcal K_t^{\mathrm{ava}}
    \triangleq
    \left\{
        k\in\mathcal K:
        Q_k[t]>0,\ t\geq\ell_k
    \right\}.
    \label{eq:eligible_set}
\end{equation}
The condition \(t\geq\ell_k\) ensures that the branch input of agent \(k\) is locally available, whereas \(Q_k[t]>0\) indicates that its payload has not been completely delivered. Therefore, \(\mathcal K_t^{\mathrm{ava}}\) contains precisely the agents that can be assigned RBs in slot \(t\). For each RB \(f\), a causal communication policy selects an agent according to
\begin{equation}
    k^{\star}(t,f)
    =
    \pi^{\mathrm C}\!\left(
        \mathcal H_t,
        f,
        \mathcal K_t^{\mathrm{ava}}
    \right),
    \label{eq:pluggable_scheduler}
\end{equation}
provided that \(\mathcal K_t^{\mathrm{ava}}\neq\varnothing\); otherwise, the RB remains unassigned. The selected agent determines the corresponding allocation indicators \(x_{k,f}[t]\) used in \eqref{eq:slot_payload}--\eqref{eq:queue_evolution}. This policy-agnostic interface requires only that \(\pi^{\mathrm C}\) satisfy the causality and resource-allocation constraints in Section~\ref{subsec:communication_model}. 

After the RB allocation for slot \(t\) has been determined, the communication module computes the delivered payload \(d_k[t]\) and updates \(Q_k[t+1]\) according to~\eqref{eq:queue_evolution}. The first transition from \(Q_k[t]>0\) to \(Q_k[t+1]=0\) marks the physical transmission completion of branch \(k\). At that instant, the module records \(a_k=t\Delta\) and emits the completion event
\begin{equation}
    e_k
    \triangleq
    \operatorname{Complete}(k,a_k).
    \label{eq:completion_event}
\end{equation}
Each event is generated exactly once and certifies that the corresponding branch input has been completely delivered. A predicted completion time may inform a causal allocation decision, but it cannot directly trigger \eqref{eq:completion_event}; event generation must be supported by an actual queue transition.

\subsection{RTCC Release Controller}
\label{subsec:release_controller}
The RTCC release controller transforms valid transmission-completion events into executable DNN branch tasks by managing the lifecycle of each communication-dependent branch. Specifically, each branch is associated with a state variable
\begin{equation}
 \hspace{-1mm}
    z_k\in
    \{\texttt{WAITING},\texttt{RELEASED},
      \texttt{RUNNING},\texttt{FINISHED}\}.
    \label{eq:branch_state}
\end{equation}
These states follow a strictly ordered lifecycle:
$\texttt{WAITING}\rightarrow\texttt{RELEASED}\rightarrow
\texttt{RUNNING}\rightarrow\texttt{FINISHED}$.
To clarify state semantics, \texttt{WAITING} denotes pending input completion, while \texttt{RELEASED} signifies a fully received branch input with the corresponding encoder enqueued for execution. Accordingly, the branch state transitions to \texttt{RUNNING} once the GPU executor dispatches the encoder and switches to \texttt{FINISHED} after encoder execution finishes and the output features are stored. The initial state transition is governed by the RTCC release controller, whereas the subsequent runtime state updates are managed by the dependency-aware GPU executor.

All branches are initialized to the \texttt{WAITING} state. Upon receiving the completion event $e_k$, the controller first validates the integrity, size, and tensor consistency of the received payload. Only after successful validation does the controller atomically update $z_k$ to \texttt{RELEASED}, set the physical release time to $r_k^{\mathrm R}=a_k$, and insert encoder $E_k$ into the ready queue ordered by the tuple $(r_k^{\mathrm R},k)$, where the superscript ${\mathrm R}$ denotes RTCC execution. Thus, the physical release time determines the ready-queue order, while the branch index provides deterministic tie-breaking for simultaneously completed branches. Duplicate, premature, malformed, or inconsistent completion events are rejected to ensure valid and deterministic scheduling.

To formalize a valid branch release, define the completion-slot index of a branch with finite arrival time as $\tau_k\triangleq a_k/\Delta\in\mathcal T$. Any legitimate state transition satisfies
\begin{equation}
    z_k\neq\texttt{WAITING}
    \Longrightarrow
    Q_k[\tau_k+1]=0
    \ \text{and}\ 
    r_k^{\mathrm R}=a_k.
    \label{eq:release_certificate}
\end{equation}
Thus, no branch can exit the \texttt{WAITING} state before physical transmission completion. Combined with the execution constraint in \eqref{eq:encoder_release}, this release certificate guarantees the practical timing relationship $s_{E_k}^{\mathrm R}\geq r_k^{\mathrm R}=a_k$. Finally, if any branch input remains undelivered at the end of the communication horizon, the current frame is marked communication-infeasible. Consequently, the shared fusion and downstream task subgraph is withheld from execution, and no complete DNN inference result is generated.

\subsection{Dependency-Preserving GPU Execution}
\label{subsec:release_aware_executor}

The executor schedules released branch encoders on a single nonpreemptive, work-conserving GPU compute stream to process sequentially arriving inputs while strictly preserving the DNN dependencies and fixed agent-indexed feature order. Specifically, an encoder becomes ready for GPU execution only after its complete branch input arrives. Accordingly, all ready encoders are placed in a queue sorted by input-completion time \(a_k\); agent indices resolve ties deterministically when multiple inputs finish simultaneously. Following this queuing discipline and the work-conserving property, the executor immediately launches the encoder at the queue head whenever the queue is nonempty and the GPU is available. Once dispatched, an encoder runs to completion without preemption.

To formalize the scheduling behavior, let \(\sigma(1),\dots,\sigma(K)\) denote the agent indices ordered by the tuple \((a_k,k)\), where \(E_{\sigma(j)}\) is the \(j\)-th encoder dispatched for execution. Let \(g_j^{\mathrm R}\) denote the time at which the GPU becomes available after processing the first \(j\) encoders under RTCC, with \(g_0^{\mathrm R}=0\). The RTCC start and finish times of \(E_{\sigma(j)}\) then satisfy
\begin{align}
    &s_{E_{\sigma(j)}}^{\mathrm R}
    =
    \max\left\{
        a_{\sigma(j)},
        g_{j-1}^{\mathrm R}
    \right\},
    \label{eq:rtcc_encoder_start}\\
    &f_{E_{\sigma(j)}}^{\mathrm R}
    =
    s_{E_{\sigma(j)}}^{\mathrm R}
    +
    c_{E_{\sigma(j)}},
    \label{eq:rtcc_encoder_finish}\\
    &g_j^{\mathrm R}
    =
    f_{E_{\sigma(j)}}^{\mathrm R},
    \qquad j \in \mathcal{K}.
    \label{eq:rtcc_gpu_availability}
\end{align}
The two terms in \eqref{eq:rtcc_encoder_start} correspond to the input-readiness and GPU-resource constraints, respectively. If the branch input is ready before the previous encoder finishes, the next encoder starts immediately at \(g_{j-1}^{\mathrm R}\). Otherwise, the GPU remains idle until the corresponding input arrives at \(a_{\sigma(j)}\). Critically, encoder scheduling relies on actual input-completion events rather than predicted arrival times, thereby guaranteeing deterministic and valid execution.

To manage the complete lifecycle of encoding tasks, the executor tracks each encoder's runtime state. When \(E_k\) is selected from the ready queue, its state transitions from \texttt{RELEASED} to \texttt{RUNNING}. Upon completion, multi-scale output features are written to storage slots assigned by the original agent index \(k\), and the encoder state switches to \texttt{FINISHED}. Because storage locations follow static agent indexing rather than execution order, variations in input arrivals and encoder scheduling do not reorder the feature tensors supplied to the fusion module. Subsequently, the shared post-join subgraph executes only after all \(K\) encoders reach the \texttt{FINISHED} state. Operators within this subgraph run in topological order, with each node starting only after all its predecessors have completed. Consequently, all downstream DNN computations, including multi-scale fusion, upsampling, tensor concatenation, and the classification and regression heads, proceed unchanged.  

\subsection{Task-Weighted MaxRate Instantiation}
\label{subsubsec:task_weighted_maxrate}
Within the causal communication interface established above, Task-Weighted MaxRate is selected as the default RB-allocation policy. This scheduler combines instantaneous uplink rates with static task-relevance weights, thereby prioritizing agents with both favorable transmission conditions and high task relevance.

To instantiate the policy-agnostic scheduling framework, each sensing agent \(k\) is assigned a fixed offline task-relevance weight \(q_k>0\), which remains unchanged during online operation. For RB \(f\) in slot \(t\), the scheduling score of an eligible agent is
\begin{equation}
    \varphi_{k,f}[t]
    \triangleq
    q_k R_{k,f}[t],
    \qquad
    k\in\mathcal K_t^{\mathrm{ava}},
    \label{eq:task_weighted_score}
\end{equation}
where \(R_{k,f}[t]\) is the instantaneously observable uplink rate. RB allocation follows
\begin{equation}
    k^{\star}(t,f)
    =
    \underset{k\in\mathcal K_t^{\mathrm{ava}}}
    {\operatorname*{arg\,max}}\;
    \varphi_{k,f}[t].
    \label{eq:task_weighted_maxrate}
\end{equation}
Accordingly, the allocation indicator satisfies
\begin{equation}
    x_{k,f}[t]
    =
    \begin{cases}
        1, & k=k^{\star}(t,f),\\
        0, & \text{otherwise},
    \end{cases}
    \label{eq:task_weighted_allocation}
\end{equation}
when \(\mathcal K_t^{\mathrm{ava}}\neq\varnothing\); if no eligible agent exists, RB \(f\) remains unallocated. Ties in \eqref{eq:task_weighted_maxrate} are resolved in favor of the smallest agent index to guarantee deterministic scheduling decisions.

When all agents have identical weights \(q_k\), the scheduler reduces to standard MaxRate scheduling. Because each decision uses only current rate measurements, static task weights, and the current eligible-agent set, the scheme satisfies the causality constraints. It requires no future channel state information, predicted completion events, or runtime GPU-state feedback. After RB assignments take effect, the residual payloads are updated according to \eqref{eq:queue_evolution}. Once a branch payload is fully delivered, a physical completion event is generated and forwarded to the RTCC controller.

Task-Weighted MaxRate is one feasible instantiation of the communication interface and imposes no inherent limitation on RTCC. Other causal scheduling algorithms can replace the rule in \eqref{eq:task_weighted_maxrate} without modifying the subsequent release controller or dependency-aware GPU executor. Its per-slot computational complexity is \(\mathcal O(FK)\), because scores are evaluated for up to \(K\) eligible agents on each of the \(F\) RBs.

\subsection{Physical CUDA Realization}
\label{subsec:cuda_realization}
Unlike an abstract delay model that merely specifies input-arrival times, the CUDA implementation enforces the established timing constraint within the physical GPU dispatch pipeline. Specifically, branch inputs remain in pinned host memory before \(a_k\), preventing premature H2D transfer and encoder execution. At \(a_k\), the communication layer emits a completion event \(e_k\) to notify the RTCC controller that the input has been fully received. Upon receiving \(e_k\), the controller launches an asynchronous H2D copy on a dedicated CUDA transfer stream. Because stream submission does not imply immediate GPU data availability, a CUDA event records the completion of each H2D transfer, and the serial compute stream waits for the corresponding event before launching encoder \(E_k\). Thus, \(e_k\) certifies communication completion, while the CUDA event certifies successful data transfer to the GPU. After execution, \(E_k\) stores its output features in memory slots statically assigned to agent \(k\). The shared fusion and downstream subgraph then executes only after all \(K\) encoders have completed their feature writes. This hardware pipeline guarantees that the practical encoder start time satisfies \(s_{E_k}^{\mathrm R}\geq a_k\).

\subsection{RTCC Algorithm and Computational Complexity}
\label{subsec:rtcc_algorithm_complexity}
Algorithm~\ref{alg:rtcc} summarizes RTCC with an arbitrary causal communication policy. Because the considered communication policies do not access GPU runtime states, the algorithm separates arrival generation and release-aware DAG execution into two logical phases. An online implementation pipelines incoming completion events with ongoing GPU execution while preserving the same release order and single-stream scheduling outcome. Initialization ensures that each branch generates at most one completion event. The main loop implements causal RB allocation and exact payload accounting, and the embedded feasibility check verifies complete delivery of all inputs required by the detection pipeline. The ordered event queue subsequently enforces the release certificate and single-stream execution discipline. Fixed-index feature storage and topological execution of the post-join subgraph jointly preserve the DNN dependencies and output mapping.

For communication policies that evaluate all \(K\) eligible agents on every RB, the per-slot computational overhead is \(\mathcal O(FK)\), yielding an overall horizon complexity of \(\mathcal O(T_{\max}FK)\). Task-Weighted MaxRate computes priority scores and performs an \(\operatorname*{arg\,max}\) selection for each RB. Because each branch generates at most one completion event, constructing the ordered release queue incurs \(\mathcal O(K\log K)\) time and \(\mathcal O(K)\) memory. Given a precomputed topological ordering, dependency tracking for one complete DNN execution costs \(\mathcal O(|\mathcal{V}|+|\mathcal{E}|)\). These scheduling and control operations contribute negligible overhead; the physical runtime is dominated by data transfers and CUDA kernels.

\begin{algorithm}[!t]
\caption{RTCC With a Causal Communication Policy}
\label{alg:rtcc}
\begin{algorithmic}[1]
\Require $\mathbf B$, $\boldsymbol\ell$, $\Delta$, $F$, and $T_{\max}$
\Require Profiled DAG $(\mathcal G,\mathbf c)$ and causal policy
$\pi^{\mathrm C}$
\Ensure $(\widehat{Y}_{\mathrm R},T_{\mathrm R},\mathbf a)$ or
\textsc{Infeasible}
\State Set $Q_k[1]\gets B_k$ and $a_k\gets+\infty$ for all $k\in\mathcal K$
\State Set $z_k\gets\texttt{WAITING}$ for all $k\in\mathcal K$
\For{$t=1,\ldots,T_{\max}$}
    \State Observe $\mathbf R[t]$ and set $\mathbf x[t]\gets\mathbf 0$
    \State Construct $\mathcal K_t^{\mathrm{ava}}$ using
    \eqref{eq:eligible_set}
    \For{$f=1,\ldots,F$}
        \If{$\mathcal K_t^{\mathrm{ava}}\neq\varnothing$}
            \State Select $k^\star(t,f)$ using \eqref{eq:pluggable_scheduler}
            \State $x_{k^\star(t,f),f}[t]\gets1$
        \EndIf
    \EndFor
    \State Compute $d_k[t]$ and update $Q_k[t+1]$ for all $k\in\mathcal K$
    \State $\mathcal C_t\gets
    \{k\in\mathcal K:Q_k[t]>0,\ Q_k[t+1]=0\}$
    \ForAll{$k\in\mathcal C_t$}
        \State $a_k\gets t\Delta$; emit
        $e_k\gets\operatorname{Complete}(k,a_k)$
    \EndFor
    \If{$a_k<+\infty$ for all $k\in\mathcal K$}
        \State \textbf{break}
    \EndIf
\EndFor
\If{$a_k=+\infty$ for any $k\in\mathcal K$}
    \State \Return \textsc{Infeasible}
\EndIf
\State Order validated events in $\mathcal Q_{\mathrm{rel}}$ by $(a_k,k)$
\State Set $g\gets0$ and initialize agent-indexed feature storage
\While{$\mathcal Q_{\mathrm{rel}}\neq\varnothing$}
    \State $(k,a_k)\gets
    \operatorname{pop}(\mathcal Q_{\mathrm{rel}})$
    \State $z_k\gets\texttt{RELEASED}$;
    $s_{E_k}^{\mathrm R}\gets\max\{a_k,g\}$
    \State $z_k\gets\texttt{RUNNING}$; execute $E_k$
    \State $g\gets s_{E_k}^{\mathrm R}+c_{E_k}$; store feature in slot $k$
    \State $z_k\gets\texttt{FINISHED}$
\EndWhile
\State Execute the shared post-join DAG in topological order
\State $T_{\mathrm R}\gets f_o^{\mathrm R}$;
$\widehat{Y}_{\mathrm R}\gets$ complete inference output
\State \Return $(\widehat{Y}_{\mathrm R},T_{\mathrm R},\mathbf a)$
\end{algorithmic}
\end{algorithm}

\section{Experimental Evaluation}
\label{sec:evaluation}

This section evaluates the performance of RTCC through physical GPU experiments and large-scale trace-driven simulations. For comparison, Barrier is adopted as the execution baseline, which releases all branch encoders only after all agent inputs have arrived. Barrier represents the conventional stage-separated execution implicit in representative edge-assisted cooperative-perception and task-offloading pipelines, where collaborative data delivery is completed before the corresponding aggregate edge inference/processing stage is executed~\cite{Liu2023AdaMap,Zaki2024QCPTO,Dong2025ITORA}. These approaches optimize what is transmitted, offloaded, or allocated, but do not expose each mandatory payload-completion event as a release signal to a specific GPU branch. Thus, Barrier is used here to isolate the execution-side waiting penalty that remains when branch-level communication completion is not coupled to GPU dispatch. First, end-to-end latency is measured on the target GPU to establish its practical performance. Second, the release-overlap mechanism is validated by comparing the measured gains with the profiled-DAG predictions. Third, robustness is examined across different communication loads and causal scheduling policies. Finally, output-invariance experiments confirm that release-triggered execution preserves the final detection results.

\subsection{Experimental Setup}
\label{subsec:experimental_setup}

The principal experimental parameters are summarized in Table~\ref{tab:experimental_configuration}. The following subsections describe the DNN workload, communication model, compared execution rules, and evaluation methodology.

\begin{table*}[!t]
\centering
\caption{Experimental Configuration}
\label{tab:experimental_configuration}
\footnotesize
\setlength{\tabcolsep}{4.5pt}
\renewcommand{\arraystretch}{1.08}
\begin{tabular}{lll|lll}
\toprule
\textbf{Category} & \textbf{Parameter} & \textbf{Setting}
& \textbf{Category} & \textbf{Parameter} & \textbf{Setting} \\
\midrule
DNN
& Detector
& PointPillars-based attentive fusion
& Communication
& Number of agents \(K\)
& 6 \\

DNN
& Inference precision
& FP32
& Communication
& Number of RBs \(F\)
& 8 \\

Platform
& GPU
& RTX 2000 Ada Laptop
& Communication
& Slot duration \(\Delta\)
& 1 ms \\

Platform
& Framework
& PyTorch 2.11.0
& Communication
& Horizon \(T_{\max}\)
& 900 ms \\

Platform
& CUDA
& 12.8
& Communication
& Service deadline
& 400 ms \\

Profiling
& Warm-up runs
& 5
& Channel
& Deployment area
& \(80\times50\) m\(^2\) \\

Profiling
& Repetitions per node
& 30
& Channel
& Path-loss exponent
& 2.15 \\

Profiling
& Encoder runtime
& 6.11--6.52 ms
& Channel
& Reference RB rate
& 250--4,500 kbit/s \\

Workload
& Branch-input size \(B_k\)
& 60.6--65.2 KiB
& Load sweep
& Payload scale \(\gamma\)
& \(\{0.5,1.0,1.5,2.0\}\) \\

GPU evaluation
& Independent traces
& 30 per load
& Load sweep
& Bandwidth scale \(\beta\)
& \(\{0.6,0.8,1.0,1.2,1.4\}\) \\

GPU evaluation
& Repetitions per trace
& 3
& Scheduling
& Causal policies
& 5 \\
\bottomrule
\end{tabular}
\end{table*}

\subsubsection{DNN Workload and GPU Platform}
\label{subsubsec:dnn_gpu_platform}
 
\begin{figure}[!t]
    \centering
    \includegraphics[width=\columnwidth]{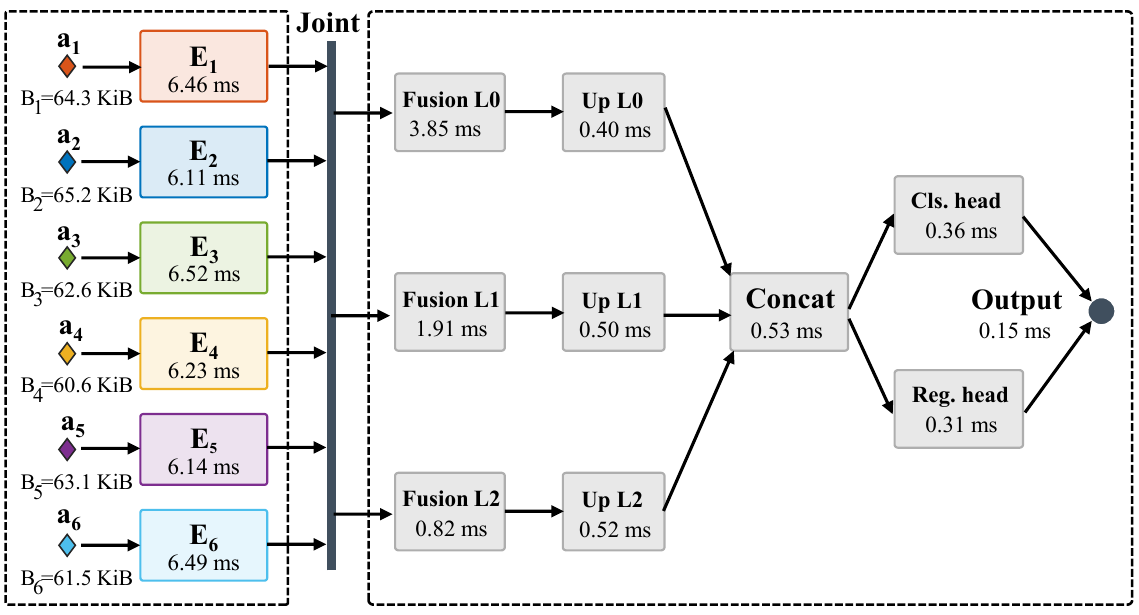}
    \caption{Profiled execution DAG of the trained DNN, with diamond nodes representing branch-input completion events and node values showing measured median GPU service times.}
    \label{fig:profiled_dag}
\end{figure}

The edge-side DNN workload is instantiated by a trained PointPillars-based intermediate-fusion detector with attentive multi-scale fusion from
OpenCOOD~\cite{Xu2022OPV2V}. For each inference frame, six agent inputs are processed by six execution instances of the shared pretrained encoder, after which the multi-scale features are fused and passed to the original classification and regression heads. FP32 inference is implemented in PyTorch~2.11.0 and CUDA~12.8 on an NVIDIA RTX~2000 Ada Generation Laptop GPU with 8~GB of memory. Fig.~\ref{fig:profiled_dag} represents the trained detector as an execution-level DAG without modifying its operators, dependencies, or weights. The DAG preserves the six independent encoder branches and their all-agent fusion dependency. After five warm-up runs, each node is measured over 30 repetitions using CUDA events, and the median runtime is used as its GPU service time. The resulting encoder service times range from 6.11 to 6.52~ms, while the six measured branch-input sizes range from 60.6 to 65.2~KiB. Consequently, only the independent encoders can overlap ongoing uplink transmission, whereas the post-join subgraph must wait for all encoder outputs.

\subsubsection{Communication Workload and Channel Configuration}
\label{subsubsec:communication_setup}

The slot-level OFDMA simulator considers \(K=6\) mobile agents and \(F=8\) RBs, with a slot duration of \(\Delta=1\)~ms, a communication horizon of
\(T_{\max}=900\)~ms, and a service deadline of 400~ms. The agents move within an \(80\times50\)-m area served by a centrally located edge server. The channel model incorporates distance-dependent attenuation with a path-loss exponent of 2.15, agent-specific blockage, slot-level small-scale variation, and RB-dependent frequency selectivity. The reference rate of each agent--RB pair is independently drawn from 250 to 4,500~kbit/s before these channel effects are applied. To vary the communication load, a payload factor \(\gamma\) scales the branch-input sizes \(B_k\), while a bandwidth factor \(\beta\) scales the instantaneous RB rates. The nominal operating point is \((\gamma,\beta)=(1.0,0.8)\), and robustness is evaluated over
\begin{equation}
    \gamma\in\{0.5,1.0,1.5,2.0\},\quad
    \beta\in\{0.6,0.8,1.0,1.2,1.4\}.
    \label{eq:communication_load_sweep}
\end{equation}

\subsubsection{Schedulers, Baseline, and Metrics}
\label{subsubsec:scheduling_baselines}

Five representative causal schedulers are considered, prioritizing instantaneous throughput, residual payload, branch completion, DAG criticality, and task relevance, respectively. For a unified representation, RB $f$ is assigned to
\begin{equation}
    k^\star(t,f)=
    \operatorname*{arg\,max}_{k\in\mathcal K_t^{\mathrm{ava}}}
    s_{k,f}[t],
    \label{eq:evaluation_schedulers}
\end{equation}
where
\begin{equation}
\hspace{-3mm}
s_{k,f}[t]=
\begin{cases}
R_{k,f}[t], & \text{MaxRate},\\
Q_k[t]R_{k,f}[t], & \text{MaxWeight},\\
-Q_k[t]/b_{k,f}[t], & \text{EarliestRelease},\\
\omega_kR_{k,f}[t], & \text{DAGRank},\\
q_kR_{k,f}[t], & \text{Task-Weighted MaxRate}.
\end{cases}
\label{eq:evaluation_scores}
\end{equation}
Here, $b_{k,f}[t]$ is the payload deliverable in the current slot, $\omega_k$ is the normalized upward rank obtained from the profiled DAG, and
$\mathbf q=[1.00,0.80,1.20,0.70,1.05,0.60]^{\mathsf T}$ is fixed throughout the evaluation. If multiple agents achieve the same priority score, the agent with the smallest index is selected to ensure deterministic allocation.

For each communication trace, Barrier and RTCC use the same branch-arrival vector and differ only in the encoder-release rule; all DNN operations, model weights, input tensors, GPU settings, and detection postprocessing remain identical. Additionally, latency is measured from the beginning of uplink transmission using two endpoints. \(T^{\mathrm{head}}\) ends when the raw classification and regression outputs become available, whereas the primary metric \(T^{\mathrm{det}}\) further includes bounding-box decoding, score thresholding, and rotated non-maximum suppression (NMS), and ends when the final detection results are produced. For \(x\in\{\mathrm{head},\mathrm{det}\}\), the paired latency gain is defined as
\begin{equation}
    \Delta T^{x}
    \triangleq
    T_{\mathrm B}^{x}-T_{\mathrm R}^{x},
    \label{eq:evaluation_gain}
\end{equation}
where a positive value indicates that RTCC outperforms Barrier. The analytical completion time $T$ corresponds to the GPU-side DNN completion endpoint $T^{\mathrm{head}}$, whereas $T^{\mathrm{det}}$ additionally includes host-side detection postprocessing and is reported as the primary deployment-level metric. To mitigate order-dependent GPU effects, the execution order of each Barrier--RTCC pair is alternated. Each independent arrival trace is replayed three times, and the three measurements are averaged to obtain one trace-level observation. Accordingly, independent traces, rather than repeated runs, are treated as the statistical units. The uncertainty of the mean latency gain is quantified using 95\% bootstrap confidence intervals obtained from 10,000 resamples of the independent trace-level observations.

\subsection{End-to-End Performance on the Physical GPU}
\label{subsec:real_gpu_results}

\begin{table*}[!t]
\centering
\caption{Trace-Level Paired Results on the Physical GPU}
\label{tab:real_gpu_main}
\footnotesize
\setlength{\tabcolsep}{7.5pt}
\renewcommand{\arraystretch}{1.12}

\begin{tabular}{@{}lccccccccc@{}}
\toprule
\textbf{Load}
&
\(\boldsymbol{(\gamma,\beta)}\)
&
\shortstack{\(\boldsymbol{A}\)\\\textbf{p50}}
&
\shortstack{\(\boldsymbol{T^{\mathrm{head}}}\)\\
\textbf{p50 B/R}}
&
\shortstack{\(\boldsymbol{T^{\mathrm{det}}}\)\\
\textbf{p50 B/R}}
&
\shortstack{\(\boldsymbol{T^{\mathrm{det}}}\)\\
\textbf{p95 B/R}}
&
\shortstack{\(\boldsymbol{\Delta T^{\mathrm{det}}}\) \textbf{Mean}\\
\textbf{[95\% CI]}}
&
\shortstack{\textbf{Mean}\\
\textbf{Reduction}}
&
\textbf{Wins}
&
\shortstack{\textbf{Deadline (\%)}\\
\textbf{B/R}} \\
\midrule

Low
&
\((0.5,1.4)\)
&
96.0
&
170.8/127.9
&
258.5/208.0
&
287.5/235.5
&
46.7 [42.0,51.2]
&
18.1\%
&
30/30
&
100/100
\\

Nominal
&
\((1.0,0.8)\)
&
294.5
&
381.7/330.1
&
459.1/408.6
&
486.3/431.1
&
48.6 [44.3,53.1]
&
10.6\%
&
30/30
&
0/30
\\

High
&
\((2.0,0.6)\)
&
598.0
&
742.2/629.1
&
801.3/694.6
&
868.5/746.5
&
99.5 [81.2,115.7]
&
12.2\%
&
28/30
&
0/0
\\

\bottomrule
\end{tabular}
\end{table*}

\begin{figure*}[!t]
    \centering
    \includegraphics[width=0.9\textwidth]{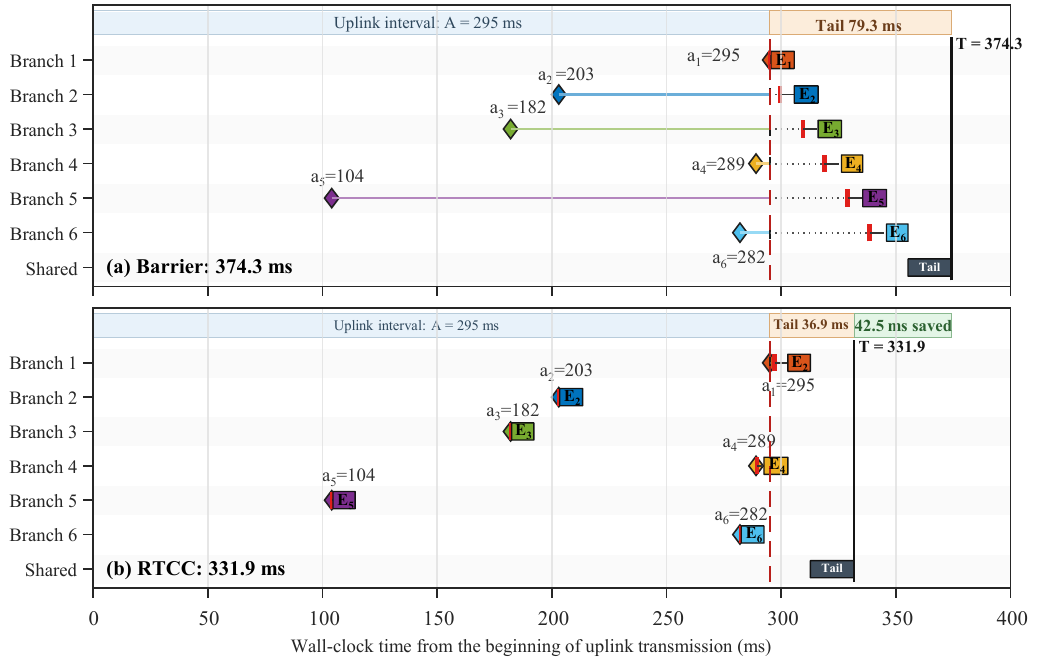}
    \caption{Representative physical CUDA timelines under the same
    branch-arrival vector. Barrier releases all encoders at the final arrival
    \(A\), whereas RTCC overlaps eligible encoder execution with the remaining
    uplink transmission.}
    \label{fig:real_cuda_timeline}
\end{figure*}

Fig.~\ref{fig:real_cuda_timeline} provides a mechanism-level illustration using one representative trace at the nominal communication load. The
timeline ends at the completion of the shared fusion and task heads, corresponding to \(T^{\mathrm{head}}\), and shows how RTCC moves eligible
encoder executions into the uplink interval. To evaluate whether this benefit generalizes beyond the illustrated trace, the physical-GPU experiment further considers low \((\gamma,\beta)=(0.5,1.4)\), nominal \((1.0,0.8)\), and high \((2.0,0.6)\) communication loads. At each load, 30 independent arrival traces are replayed three times, yielding 270 counterbalanced Barrier--RTCC pairs in total. The repetitions are averaged within each trace, resulting in 30 independent trace-level observations per load. Table~\ref{tab:real_gpu_main} summarizes both the task-head latency
\(T^{\mathrm{head}}\) and the complete-detection latency \(T^{\mathrm{det}}\). For the primary complete-detection endpoint, RTCC reduces
the median latency from 258.5 to 208.0~ms at low load, from 459.1 to 408.6~ms at nominal load, and from 801.3 to 694.6~ms at high load. The
corresponding mean paired gains are 46.7, 48.6, and 99.5~ms, with RTCC improving 30/30, 30/30, and 28/30 independent traces, respectively.

\begin{figure*}[!t]
    \centering
    \includegraphics[width=\textwidth]{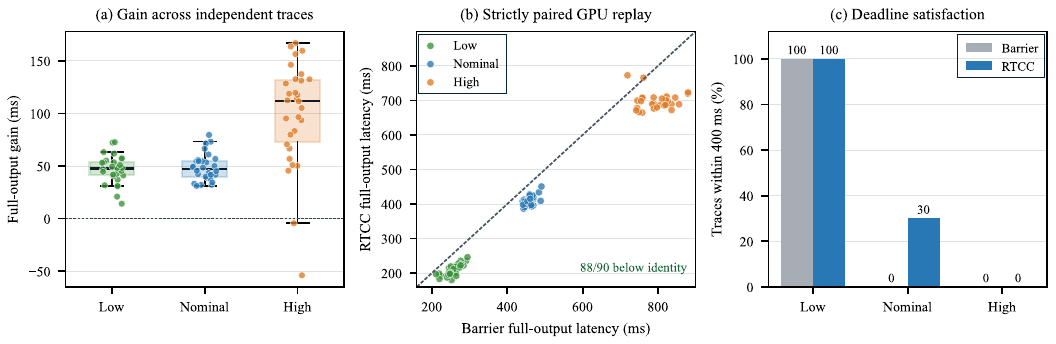}
    \caption{Trace-level physical-GPU results. (a) Paired wall-clock reductions
    under three communication loads, where circles denote trace-level
    observations and horizontal lines within the boxes denote medians.
    (b) Paired Barrier and RTCC completion times; the dashed line denotes
    equal latency. (c) Percentage of traces meeting the 400-ms service
    deadline.}
    \label{fig:real_gpu_gain}
\end{figure*}

Fig.~\ref{fig:real_gpu_gain} summarizes the trace-level physical-GPU results from three complementary perspectives: the distribution of paired latency gains, the direct latency comparison between Barrier and RTCC, and service-deadline satisfaction. Panel (a) retains all trace-level observations and presents the gain distributions under the three communication loads. RTCC improves all 30 low-load and all 30 nominal-load traces, as well as 28 of the 30 high-load traces. The two high-load regressions are associated with host-side postprocessing and Windows/WDDM wall-clock variation.
Panel (b) directly compares the paired Barrier and RTCC completion times, with 88 of the 90 points lying below the identity line. Across all loads, the mean complete-detection reduction is 64.96~ms (95\% CI: 57.11--72.81~ms). The task-head endpoint improves on 89 of 90 traces and has a mean reduction of 64.38~ms (95\% CI: 57.05--72.05~ms). These paired results demonstrate a distributional improvement, rather than deterministic speedup on every host-timed run. Panel (c) reports satisfaction of the 400-ms service deadline. Both modes satisfy the deadline on all low-load traces. At the nominal point, complete-detection deadline satisfaction rises from 0\% under Barrier to 30\% under RTCC; neither mode meets the deadline under the deliberately communication-heavy high-load setting.

\subsection{Mechanism Validation and Load Robustness}
\label{subsec:mechanism_robustness}

\begin{figure}[!t]
    \centering
    \includegraphics[width=\columnwidth]
    {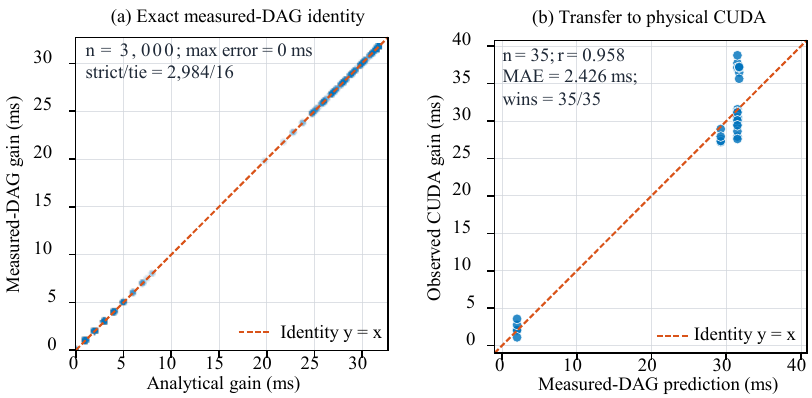}
    \caption{Mechanism validation at two evidence layers. (a) Exact agreement between the analytical pre-arrival encoder work and the measured-DAG gain over 3,000 paired trials. (b) Agreement between the measured-DAG prediction and the observed wall-clock gain over 35 resident-input CUDA pairs. The dashed line denotes identity.}
    \label{fig:mechanism_validation}
\end{figure}

Fig.~\ref{fig:mechanism_validation} illustrates the release-overlap mechanism in terms of analytical consistency and physical-GPU transferability. Panel (a) validates the analytical gain identity via 3,000 paired measured-DAG trials covering four payload scales, five bandwidth scales, five causal schedulers, and 30 channel traces. Each trial pairs Barrier and RTCC under identical branch-arrival vectors and profiled GPU service times, with only the encoder-release rule differing. The horizontal axis denotes the pre-arrival encoder work \(W_{\mathrm{pre}}\), and the vertical axis denotes the measured-DAG latency gain \(T_{\mathrm{B}}^{\mathrm{DAG}}-T_{\mathrm{R}}^{\mathrm{DAG}}\). All 3,000 samples exactly follow the identity line, strictly confirming \(T_{\mathrm{B}}^{\mathrm{DAG}}-T_{\mathrm{R}}^{\mathrm{DAG}} = W_{\mathrm{pre}}\). RTCC yields positive gains in 2,984 trials, while the remaining 16 cases are zero-overlap boundaries with no performance regression. Panel (b) further verifies the mechanism's practical validity using 35 paired resident-input CUDA runs on the physical GPU, where the axes represent DAG-predicted gains and measured wall-clock gains, respectively. The two sets of results agree closely with \(r=0.958\) and a mean absolute error of 2.426~ms, with all tested cases yielding positive gains. This demonstrates that the profiled DAG faithfully characterizes the dominant release-overlap effect under controlled GPU execution. In comparison, the online-staging experiments in Section~\ref{subsec:real_gpu_results} involve host overhead, system jitter, and GPU power-state fluctuations, thereby providing practical deployment evidence rather than exact analytical verification.

\begin{figure*}[t]
    \centering
    \includegraphics[width=0.98\textwidth]
    {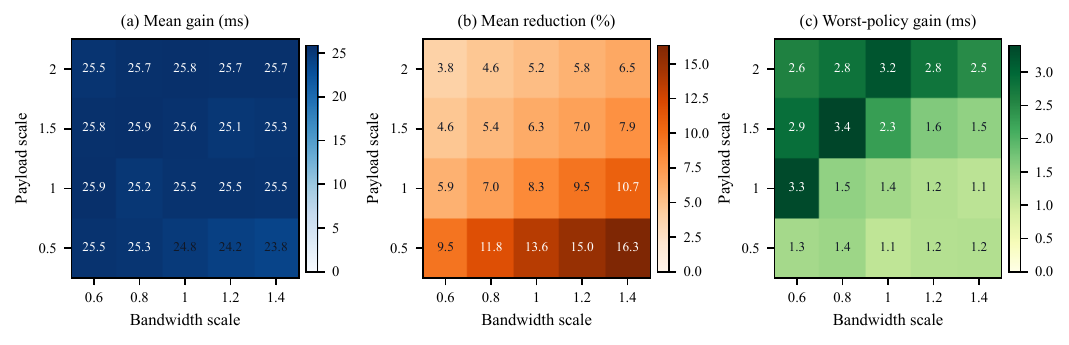}
    \caption{Measured-DAG robustness over payload and bandwidth scales. (a) Mean absolute gain across five causal communication policies. (b) Mean relative latency reduction. (c) Mean gain of the worst policy in each load cell. Each scheduler--load cell contains 30 paired trials.}
    \label{fig:load_robustness}
\end{figure*}

Fig.~\ref{fig:load_robustness} summarizes the complete 3,000-pair sweep. The mean absolute reduction across the five schedulers remains between 23.8 and 25.9~ms over the 20 payload--bandwidth cells. The relative reduction ranges from 3.8\% under the most communication-dominated setting to 16.3\% when the communication load is light. Even the worst scheduler in each cell retains a positive mean gain of 1.1--3.4~ms. Moreover, all 100 scheduler--load cells have a positive mean gain and a positive lower bound of the paired bootstrap confidence interval. These results show that RTCC removes unnecessary computation waiting across a broad range of communication conditions, although its percentage benefit decreases when wireless transmission dominates the end-to-end latency.

\subsection{Compatibility with Causal Communication Policies}
\label{subsec:scheduler_compatibility}

\begin{figure}[t]
    \centering
    \includegraphics[width=0.48\textwidth]
    {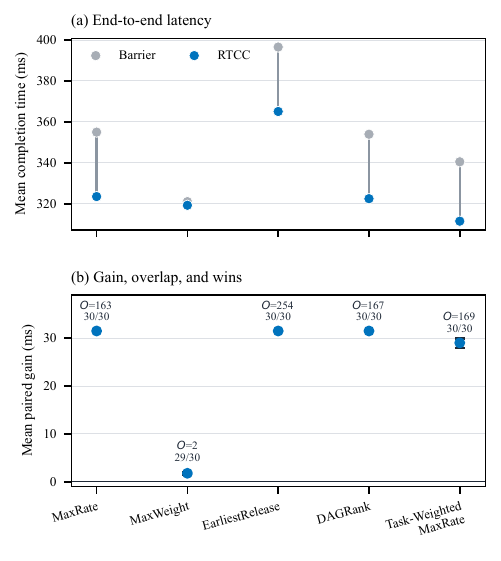}
    \caption{Compatibility with causal communication policies at the nominal load. (a) Mean Barrier and RTCC completion times. (b) Mean paired gain with a 95\% confidence interval; annotations give the mean overlap window $O$ in milliseconds and RTCC wins.}
    \label{fig:scheduler_compatibility}
\end{figure}

Fig.~\ref{fig:scheduler_compatibility} separates the communication policy from the release architecture at the nominal load. RTCC lowers the mean completion time for every scheduler. The reductions are 31.49~ms for MaxRate, DAGRank, and EarliestRelease, 29.01~ms for Task-Weighted MaxRate, and 1.77~ms for MaxWeight. The small MaxWeight gain is not a failure of the release controller: its mean usable overlap window is only 1.77~ms, compared with 163--254~ms for the other policies. MaxWeight improves in 29 of 30 pairs, whereas each other policy improves in all 30 pairs. The scheduler producing the largest RTCC release gain is not necessarily the one producing the smallest total latency. Task-Weighted MaxRate attains a mean RTCC completion time of 311.45~ms, which is 7.77~ms lower than the strongest fixed RTCC alternative, MaxWeight, despite the latter having the smallest Barrier completion time. Across the 20 load cells, the best RTCC communication policy changes: Task-Weighted MaxRate, MaxWeight, DAGRank, and MaxRate are best in seven, six, six, and one cells, respectively. Accordingly, the contribution is the scheduler-compatible RTCC execution interface, rather than universal optimality of a fixed wireless priority rule.

\subsection{Release Fidelity and Runtime Overhead}
\label{subsec:release_overhead}

\begin{figure}[t]
    \centering
    \includegraphics[width=0.48\textwidth]
    {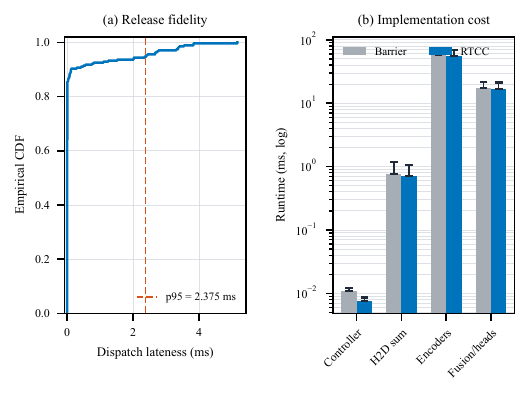}
    \caption{Physical release fidelity and implementation cost.
    (a) Empirical CDF of per-branch dispatch lateness.
    (b) Median Barrier and RTCC implementation components with 95th-percentile
    upper whiskers. The logarithmic axis exposes the microsecond-scale
    controller cost alongside millisecond-scale GPU operations.}
    \label{fig:release_overhead}
\end{figure}

Fig.~\ref{fig:release_overhead} jointly evaluates whether RTCC follows the intended causal release timing and whether its control mechanism introduces material runtime overhead. Fig.~\ref{fig:release_overhead}(a) measures the delay between the modeled release time and physical dispatch of the corresponding H2D operation. The per-branch 95th-percentile dispatch lateness is 2.375~ms. When aggregated by run, the mean, median, and 95th percentile of the six-branch average lateness are 0.224, 0.072, and 0.630~ms, respectively; the largest per-run branch delay
observed over the 45 nominal-load pairs is 5.159~ms. These measurements confirm that the physical path follows the causal release events rather than launching inputs speculatively before their communication completion. For six branches, release-plan validation, deadline construction, and stable ordering by $(a_k,k)$ require 7.63~$\mu$s per decision at the median. As shown in Fig.~\ref{fig:release_overhead}(b), this control cost is negligible relative to the GPU service times. The RTCC medians for total H2D staging, the six encoders, and the fusion/task subgraph are 0.704, 55.98, and 16.76~ms, respectively; the corresponding Barrier medians are 0.773, 57.28, and 17.18~ms. The close component times confirm that RTCC does not reduce the DNN workload. Its end-to-end gain arises from moving necessary encoder work into the communication interval.

\subsection{Output Invariance and Evaluation Scope}
\label{subsec:output_invariance}

The output-invariance experiment verifies that RTCC changes only the DNN execution timeline without altering the resulting detection output. The evaluation uses 100 labeled frames from five official OPV2V scenes, with 20 frames per scene and three collaborating agents per frame. These frames contain 1,906 ground-truth boxes and produce 1,987 final predictions. Barrier and RTCC use the same trained checkpoint, input tensors, score threshold, rotated non-maximum suppression, and average-precision implementation; they differ only in the encoder-release rule.

Barrier and RTCC achieve identical average precision (AP) values of 0.8769, 0.8698, and 0.7583 at IoU thresholds of 0.3, 0.5, and 0.7, respectively. Beyond these aggregate metrics, all 100 paired frames produce exactly identical raw classification outputs, regression outputs, final bounding boxes, confidence scores, and true-positive/false-positive sequences. These results confirm that RTCC preserves the tensor order and complete inference mapping of the trained DNN. Hence, its latency reduction arises solely from advancing the execution of eligible encoders, rather than from branch dropping, output approximation, or modification of the detection model.

The above results should be interpreted within the scope of the experimental methodology. The physical experiments execute the complete trained detector, pinned-memory input staging, CUDA event dependencies, and task heads on the target GPU, whereas the wireless completion times are generated through causal trace-driven communication emulation rather than over-the-air radio experiments. Accordingly, the results demonstrate physical GPU execution under controlled trace-driven communication-arrival patterns, rather than an end-to-end over-the-air wireless deployment. Moreover, Windows/WDDM host scheduling and GPU power-state transitions can introduce wall-clock variation, particularly under heavy communication loads; paired trace-level observations, confidence intervals, medians, and win counts are therefore reported together. Finally, the 100-frame study verifies output invariance across multiple scenes and timing patterns, rather than serving as a full-dataset detection-accuracy benchmark. The physical measurements are obtained on one target GPU platform; therefore, the reported absolute gains should not be interpreted as hardware-independent speedups. Their magnitude depends on the communication-arrival gaps, encoder service times, and platform-specific transfer and execution behavior, while the experiments establish the feasibility of the completion-triggered release path on the evaluated GPU.

\section{Conclusion}
\label{sec:conclusion}

This paper investigated the coupling between wireless input completion and GPU execution in centralized multi-agent cooperative perception.  RTCC releases each encoder branch as soon as its complete input becomes available, allowing mandatory computation to overlap ongoing uplink transmission while preserving the original fork--join dependencies and inference mapping. A completion-driven pinned-memory H2D and CUDA-event execution path was implemented on the target GPU and integrated with causal wireless schedulers through a policy-agnostic interface. Physical-GPU experiments and measured-DAG studies demonstrated consistent latency reductions across communication loads and scheduling policies, with the analytical overlap mechanism closely matching measured execution. Furthermore, output-invariance tests  confirmed that the latency gain is obtained solely by advancing eligible computation rather than modifying the perception model or its outputs. These results establish completion-triggered branch execution as a practical mechanism for reducing end-to-end latency in edge-assisted cooperative perception.

\ifCLASSOPTIONcaptionsoff
  \newpage
\fi

\end{document}